\documentclass[aps, prl, floatfix, nofootinbib, superscriptaddress, twocolumn]{revtex4-2}

\usepackage{latexsym}
\usepackage{amsmath}
\usepackage{amssymb}
\usepackage{amsfonts}

\usepackage[mathscr,scaled=1.15]{urwchancal}
\DeclareFontFamily{OT1}{pzc}{}
\DeclareFontShape{OT1}{pzc}{m}{it}%
{<-> s * [1.15] pzcmi7t}{}
\DeclareMathAlphabet{\mathpzc}{OT1}{pzc}{m}{it}

\usepackage{CJKutf8}
\usepackage{color}
\usepackage{slashed}
\usepackage{dsfont}

\usepackage{supertabular}
\usepackage{placeins}
\usepackage{epsfig}
\usepackage{graphicx}

\definecolor{purple}{rgb}{0.5,0,0.5}
\definecolor{blue}{rgb}{0.0,0,0.9}
\definecolor{prdblue}{rgb}{0.133,0.118,0.498}
\usepackage[colorlinks=true, pdfstartview=FitV, linkcolor=prdblue, citecolor= prdblue, urlcolor=prdblue]{hyperref}

\begin{document}
\begin{CJK*}{UTF8}{gbsn}

\title{$\,$\\[-6ex]\hspace*{\fill}{\normalsize{\sf\emph{Preprint no}.\
NJU-INP 091/24}}\\[1ex]
Nucleon Gravitational Form Factors}

\author{Z.-Q.\ Yao (姚照千)%
       $\,^{\href{https://orcid.org/0000-0002-9621-6994}{\textcolor[rgb]{0.00,1.00,0.00}{\sf ID}},}$}
\affiliation{School of Physics, Nanjing University, Nanjing, Jiangsu 210093, China}
\affiliation{Institute for Nonperturbative Physics, Nanjing University, Nanjing, Jiangsu 210093, China}
\affiliation{European Centre for Theoretical Studies in Nuclear Physics
            and Related Areas, \\\hspace*{1ex}Villa Tambosi, Strada delle Tabarelle 286, I-38123 Villazzano (TN), Italy}

\author{Y.-Z.\ Xu 
\makebox[3.3em][l]{\hspace*{0em}\raisebox{-0.7ex}{\includegraphics[width=3.5em]{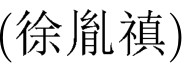}}}\;
       $^{\href{https://orcid.org/0000-0003-1623-3004}{\textcolor[rgb]{0.00,1.00,0.00}{\sf ID}}}$}
\affiliation{Dpto.~Ciencias Integradas, Centro de Estudios Avanzados en Fis., Mat. y Comp., Fac.~Ciencias Experimentales, Universidad de Huelva, Huelva 21071, Spain}
\affiliation{Dpto.~Sistemas F\'isicos, Qu\'imicos y Naturales, Univ.\ Pablo de Olavide, E-41013 Sevilla, Spain}

\author{D.\ Binosi%
    $\,^{\href{https://orcid.org/0000-0003-1742-4689}{\textcolor[rgb]{0.00,1.00,0.00}{\sf ID}}}$}
\affiliation{European Centre for Theoretical Studies in Nuclear Physics
            and Related Areas, \\\hspace*{1ex}Villa Tambosi, Strada delle Tabarelle 286, I-38123 Villazzano (TN), Italy}

\author{Z.-F.\ Cui (崔著钫)%
       $^{\href{https://orcid.org/0000-0003-3890-0242}{\textcolor[rgb]{0.00,1.00,0.00}{\sf ID}},}$}
\affiliation{School of Physics, Nanjing University, Nanjing, Jiangsu 210093, China}
\affiliation{Institute for Nonperturbative Physics, Nanjing University, Nanjing, Jiangsu 210093, China}

\author{\mbox{M.\ Ding (丁明慧)%
    $\,^{\href{https://orcid.org/0000-0002-3690-1690}{\textcolor[rgb]{0.00,1.00,0.00}{\sf ID}}}$}}
\affiliation{School of Physics, Nanjing University, Nanjing, Jiangsu 210093, China}
\affiliation{Institute for Nonperturbative Physics, Nanjing University, Nanjing, Jiangsu 210093, China}
\affiliation{Helmholtz-Zentrum Dresden-Rossendorf, Bautzner Landstra{\ss}e 400, D-01328 Dresden, Germany}

\author{K.~Raya%
       $^{\href{https://orcid.org/0000-0001-8225-5821}{\textcolor[rgb]{0.00,1.00,0.00}{\sf ID}},}$}
\affiliation{Dpto.~Ciencias Integradas, Centro de Estudios Avanzados en Fis., Mat. y Comp., Fac.~Ciencias Experimentales, Universidad de Huelva, Huelva 21071, Spain}

\author{C.\,D.\ Roberts%
       $^{\href{https://orcid.org/0000-0002-2937-1361}{\textcolor[rgb]{0.00,1.00,0.00}{\sf ID}},}$}
\affiliation{School of Physics, Nanjing University, Nanjing, Jiangsu 210093, China}
\affiliation{Institute for Nonperturbative Physics, Nanjing University, Nanjing, Jiangsu 210093, China}

\author{J.\ Rodr\'iguez-Quintero%
       $^{\href{https://orcid.org/0000-0002-1651-5717}{\textcolor[rgb]{0.00,1.00,0.00}{\sf ID}},}$}
\affiliation{Dpto.~Ciencias Integradas, Centro de Estudios Avanzados en Fis., Mat. y Comp., Fac.~Ciencias Experimentales, Universidad de Huelva, Huelva 21071, Spain}

\author{S.\,M.\ Schmidt%
    $\,^{\href{https://orcid.org/0000-0002-8947-1532}{\textcolor[rgb]{0.00,1.00,0.00}{\sf ID}},}$}
\affiliation{Helmholtz-Zentrum Dresden-Rossendorf, Bautzner Landstra{\ss}e 400, D-01328 Dresden, Germany}
\affiliation{Technische Universi\"at Dresden, 01062 Dresden, Germany}

\date{2024 October 06}


\begin{abstract}
\vspace*{-3ex}
\hspace*{-4em}\parbox[c]{\textwidth}{
\href{mailto:binosi@ectstar.eu}{binosi@ectstar.eu} (DB);
\href{mailto:phycui@nju.edu.cn}{phycui@nju.edu.cn} (ZFC);
\href{mailto:mhding@nju.edu.cn}{mhding@nju.edu.cn} (MD);
\href{mailto:cdroberts@nju.edu.cn}{cdroberts@nju.edu.cn} (CDR)}
\\[1ex]
A symmetry-preserving analysis of strong interaction quantum field equations is used to complete a unified treatment of pion, kaon, nucleon electromagnetic and gravitational form factors.
Findings include
a demonstration that the pion near-core pressure is roughly twice that in the proton, so both are significantly greater than that of a neutron star;
parton species separations of the nucleon's three gravitational form factors, in which, \emph{inter alia}, the glue-to-quark ratio for each form factor is seen to take the same constant value, independent of momentum transfer;
and
a determination of proton radii orderings, with the mechanical (normal force) radius being less than the mass-energy radius, which is less than the proton charge radius.
This body of predictions should prove useful in an era of anticipated experiments that will enable them to be tested.
\end{abstract}

\maketitle
\end{CJK*}


\noindent\emph{1.\,Introduction}\,---\,%
The nucleon mass, $m_N$, is a defining scale in Nature.  In fact, one understands the origin of almost all mass that is visible in the Universe if one grasps the source of $m_N$.  In modern high-energy physics, approximately $98$\% of $m_N$ is expected to be generated by Standard Model strong interactions, \emph{i.e}., to emerge from quantum chromodynamics (QCD) \cite{Roberts:2021nhw, Binosi:2022djx, Ding:2022ows, Ferreira:2023fva, Raya:2024ejx}.  The small remainder owes to Higgs boson couplings into QCD.

These statements are a succinct expression of an emergent hadron mass (EHM) paradigm, developed via insightful use of continuum Schwinger function methods (CSMs) -- see Ref.\,\cite{Roberts:1994dr} and citations thereof.  The three pillars of EHM are
appearance of a gluon mass scale \cite{Cornwall:1981zr},
infrared saturation and cessation of running in QCD's effective charge \cite{Binosi:2016nme, Cui:2019dwv},
and dynamical chiral symmetry breaking expressed in a nonzero chiral-limit running quark mass \cite{Lane:1974he, Politzer:1976tv}.
This paradigm is drawing support from results obtained using realistic simulations of lattice-regularised QCD (lQCD) \cite{Roberts:2021nhw, Binosi:2022djx, Ding:2022ows, Ferreira:2023fva, Raya:2024ejx}
and being/will be tested by comparisons between EHM-based predictions
and data from ongoing/anticipated experiments \cite{BESIII:2020nme, Anderle:2021wcy, Arrington:2021biu, Quintans:2022utc, Carman:2023zke, Mokeev:2023zhq}.

Regarding pion, kaon, and proton electromagnetic form factors, comparisons with data are already possible.  They support the EHM picture \cite{Yao:2024drm, Yao:2024uej}.  Additional support is provided by comparisons between CSM predictions of pion gravitational form factors \cite{Xu:2023izo} and recent lQCD results \cite{Hackett:2023nkr} -- see Ref.\,\cite[Fig.\,12]{Raya:2024ejx}.  Validated predictions for nucleon gravitational form factors would add much to the accumulating store of successes, especially if both the complete form factors and their parton decompositions were confirmed.  In these contexts, some empirical information may already be available \cite{Burkert:2018bqq, Kumericki:2019ddg, Moutarde:2019tqa}.  Moreover, lQCD has delivered results at a simulation pion mass of $\approx 0.17\,$GeV \cite{Hackett:2023rif}.

\smallskip

\noindent\emph{2.\,Nucleon Gravitational Current}\,---\,%
The proton (nucleon) has three gravitational form factors, which express all dynamical information that can be gleaned from its interaction with a $J^{PC}=2^{++}$ probe.  They are defined via the current that describes the associated scattering process, which may be written thus:
\begin{align}
m_N \Lambda_{\mu\nu}^{Ng}(Q) & = - \Lambda_+(p_f)
[
K_\mu K_\nu A(Q^2) \nonumber \\
&
\quad + i K_{\left\{\mu\right.}\!\sigma_{\left.\nu\right\}}\,\!_\rho Q_\rho J(Q^2) \nonumber \\
&
\quad + \tfrac{1}{4} (Q_\mu Q_\nu - \delta_{\mu\nu} Q^2) D(Q^2)
]
\Lambda_+(p_i)  \,,
\label{EMTproton}
\end{align}
where
$p_{i,f}$ are the momenta of the incoming/outgoing nucleon, $p_{i,f}^2=-m_N^2$,
$K=(p_i+p_f)/2$, $Q=p_f-p_i$;
all Dirac matrices are standard \cite[Sec.\,2]{Roberts:2000aa}, with $\sigma_{\mu\nu}= (i/2)[\gamma_\mu,\gamma_\nu]$;
$\Lambda_+$ is the projection operator that delivers a positive energy nucleon;
and $a_{\left\{\mu\right.}\!b_{\left.\nu\right\}}=(a_\mu b_\nu + a_\nu b_\mu)/2$.

In Eq.\,\eqref{EMTproton} \cite{Polyakov:2018zvc}:
$A$ is the nucleon mass distribution form factor;
$J$ relates to the nucleon spin distribution;
and $D$ provides information on in-nucleon pressure and shear forces.
In the forward limit, $Q^2=0$, symmetries entail $A(0)=1$, $J(0)=1/2$.
$D(0)$ is also a conserved charge, but like the axial charge, $g_A$, its value is a dynamical property of the nucleon.  It has been described as the ``last unknown global property'' of the nucleon \cite{Polyakov:2018zvc}; hence, a robust prediction of $D(0)$ is critical.

\smallskip

\noindent\emph{3.\,Continuum Calculation of Form Factors}\,---\,%
Hereafter we describe a calculation of nucleon gravitational form factors using the same approach employed for pion, kaon, and nucleon electromagnetic form factors \cite{Yao:2024drm, Yao:2024uej} and pion and kaon gravitational form factors \cite{Xu:2023izo}.
Namely, we work at leading-order in a symmetry-preserving, systematically-improvable truncation of all quantum field equations which appear in the seven-point Schwinger function that corresponds to the current in Eq.\,\eqref{EMTproton}.  This is the rainbow-ladder (RL) truncation.
After almost thirty years of use \cite{Munczek:1994zz, Bender:1996bb}, it is known to be quantitatively reliable for pion, kaon, and nucleon observables: (\emph{a}) practically, owing to widespread, successful applications \cite{Roberts:2021nhw, Ding:2022ows, Raya:2024ejx}; and (\emph{b}) because improvement schemes are available and have been tested, showing that the cumulative effect of improvement to RL truncation in these channels can be absorbed into a modest modification of the quark + quark scattering kernel \cite{Fischer:2009jm, Chang:2009zb, Chang:2013pq, Binosi:2014aea, Williams:2015cvx, Binosi:2016rxz, Binosi:2016wcx, Qin:2020jig, Xu:2022kng}.
Importantly, where reasonable comparisons are possible, contemporary CSM predictions and lQCD results are mutually consistent -- see, \emph{e.g}., Refs.\,\cite{Roberts:2021nhw, Binosi:2022djx, Ding:2022ows, Ferreira:2023fva, Raya:2024ejx, Chen:2021guo, Chang:2021utv, Lu:2023yna, Yu:2024qsd, Chen:2023zhh, Alexandrou:2024zvn}.

One might ask: What is the ``small parameter'' that explains the success of RL truncation?
The answer is straightforward.
In all systems for which nonperturbative EHM-generated feedback is small, \emph{viz}.\ ground-state channels \cite{Chang:2009zb, Qin:2020jig, Xu:2022kng}, wherein corrections may be considered diagram by diagram, cancellations can algebraically be demonstrated between new terms at each given order, with the remainder being small because:
(\emph{a}) QCD's effective charge is bounded above by $\pi$ at infrared momenta and falls monotonically from its maximum with increasing values of its spacelike argument \cite{Cui:2019dwv, Deur:2023dzc, Brodsky:2024zev}; and (\emph{b}) this ensures phase-space suppression is an effective mechanism for damping correction contributions.

In RL truncation, the nucleon wave function is obtained by solving the Faddeev equation \cite{Eichmann:2009qa} (reproduced in Fig.\,S.4 -- supplemental material, SupM).
The key element is the quark + quark scattering kernel \cite{Maris:1997tm}:
{\allowdisplaybreaks
\begin{align}
\mathscr{K}_{tu}^{rs}(k) & =\tilde{\mathpzc G}(y) T_{\mu\nu}(k) [i\gamma_\mu\frac{\lambda^{a}}{2} ]_{ts} [i\gamma_\nu\frac{\lambda^{a}}{2} ]_{ur}\,,
\label{EqRLInteraction}
%
\end{align}
$k^2 T_{\mu\nu}(k) = k^2 \delta_{\mu\nu} - k_\mu k_\nu$,  $y=k^2$.  The tensor structure specifies Landau gauge, used because it is a fixed point of the renormalisation group and that gauge for which corrections to RL truncation are minimised \cite{Bashir:2009fv}.
In Eq.\,\eqref{EqRLInteraction}, $r,s,t,u$ represent colour and spinor matrix indices.
}

Working from studies of QCD's gauge sector, one arrives at the following practicable completion of the scattering kernel \cite{Qin:2011dd, Binosi:2014aea}:
\begin{align}
\label{defcalG}
 \tilde{\mathpzc G}(y) & =
 \frac{8\pi^2}{\omega^4} D e^{-y/\omega^2} + \frac{8\pi^2 \gamma_m \mathcal{F}(y)}{\ln\big[ \tau+(1+y/\Lambda_{\rm QCD}^2)^2 \big]}\,,
\end{align}
where $\gamma_m=12/25$, $\Lambda_{\rm QCD} = 0.234\,$GeV, $\tau={\rm e}^2-1$, and ${\cal F}(y) = \{1 - \exp(-y/\Lambda_{\mathpzc I}^2)\}/y$, $\Lambda_{\mathpzc I}=1\,$GeV.
%
Analyses of gauge-sector dynamics and numerous hadron properties indicate $\omega = 0.8\,$GeV, $\omega D = 0.8\,{\rm GeV}^3$ \cite{Xu:2022kng, Xu:2023izo}.
When the product $\omega D$ is kept fixed, physical observables relevant herein remain practically unchanged under $\omega \to (1\pm 0.2)\omega$ \cite{Qin:2020rad}; thus, $\omega$ is the only degree of freedom.
(The gauge-sector context of Eq.\,\eqref{defcalG} is drawn in SupM-{\sf B}.)

Numerical methods for solving sets of coupled gap and Faddeev equations are described elsewhere -- see, \emph{e.g}., Refs.\,\cite{Maris:1997tm, Maris:2005tt, Krassnigg:2009gd, Qin:2018dqp}.  Exploiting these schemes, one may solve all equations relevant to calculation of the current in Eq.\,\eqref{EMTproton} and thereby arrive at predictions for the nucleon gravitational form factors.

As in Refs.\,\cite{Yao:2024drm, Yao:2024uej}, using Eq.\,\eqref{defcalG} with renormalisation point invariant quark current mass $\hat m_u = \hat m_d = 6.04\,$MeV, which corresponds to a one-loop mass at $\zeta=\zeta_2:=2\,$GeV of $4.19\,$MeV, the following predictions are obtained: pion mass $m_\pi = 0.14\,$GeV; nucleon mass $m_N=0.94\,$GeV; and pion leptonic decay constant $f_\pi=0.094\,$GeV.  These values align with experiment \cite{ParticleDataGroup:2024cfk}.
%

The scattering kernel involves one parameter and there is a single quark current-mass.  Both quantities are now fixed: the interaction parameter is constrained by gauge sector studies (see SupM-{\sf B}) and the quark current mass delivers a realistic pion mass.  Consequently, as in Refs.\,\cite{Yao:2024drm, Yao:2024uej, Xu:2023izo}, all calculations herein are parameter-free.

\smallskip

\noindent\emph{4.\,Gravitational Form Factors}\,---\,%
With the Poincar\'e-covariant nucleon wave function in hand, the nucleon gravitational form factors may be obtained from an interaction current of the form explained in Ref.\,\cite{Eichmann:2011vu}.
(See SupM-{\sf C} for a recapitulation.)
That current involves the graviton + quark vertex, $\Gamma_{\mu\nu}^g$, formerly unknown.
This changed with Ref.\,\cite{Xu:2023izo}, which delivered a practicable solution of the defining RL equation for $\Gamma_{\mu\nu}^g$.
Amongst other things, the study highlighted that, at low momentum transfers, graviton + hadron couplings are sensitive to contributions from the lightest tensor and scalar mesons in relevant channels, just as hadron electromagnetic form factors at low momenta are sensitive to the properties of the lightest neutral vector meson in a given quark + antiquark channel.
We follow Ref.\,\cite{Xu:2023izo} in calculating $\Gamma_{\mu\nu}^g$.
(The result is sketched in SupM-{\sf D}.)

\begin{figure}[t]
\vspace*{1.0em}

\leftline{\hspace*{0.5em}{\large{\textsf{A}}}}
\vspace*{-3.5ex}
\includegraphics[clip, width=0.405\textwidth]{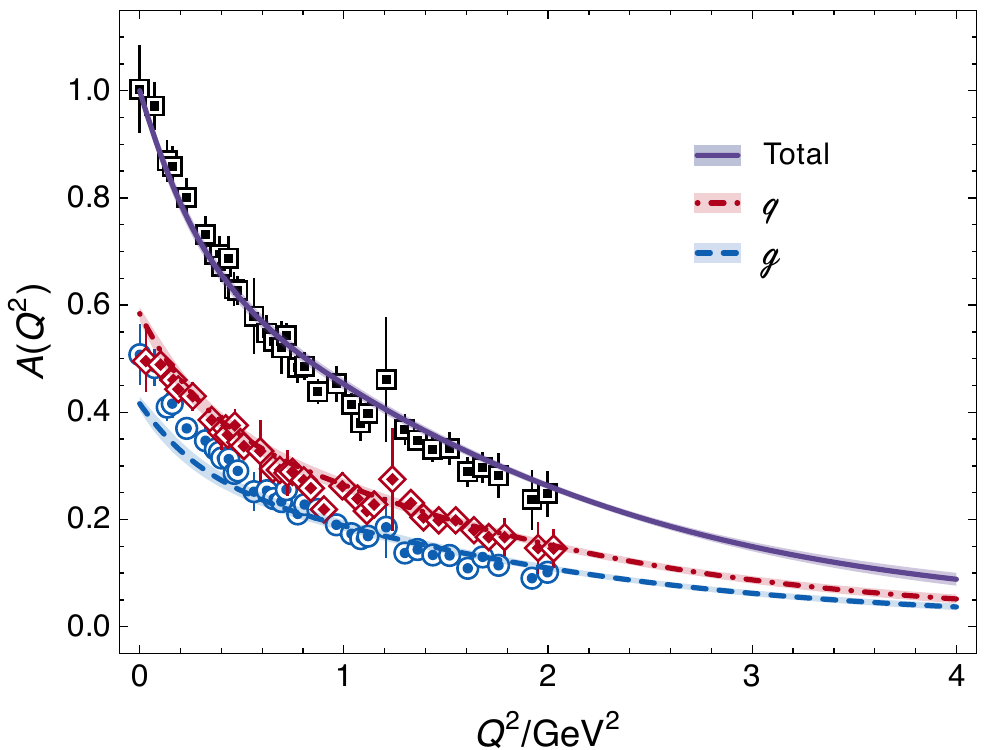}
\vspace*{1ex}
\leftline{\hspace*{0.5em}{\large{\textsf{B}}}}
\vspace*{-5ex}

\includegraphics[clip, width=0.405\textwidth]{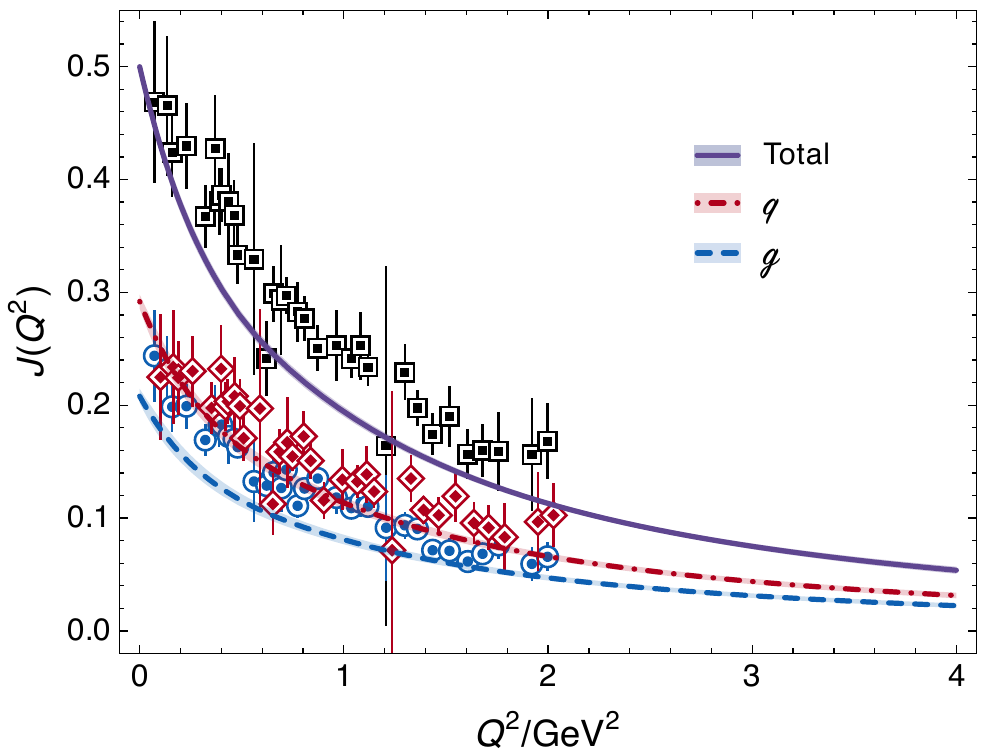}
\vspace*{3ex}
\leftline{\hspace*{0.5em}{\large{\textsf{C}}}}
\vspace*{-7ex}

\includegraphics[clip, width=0.405\textwidth]{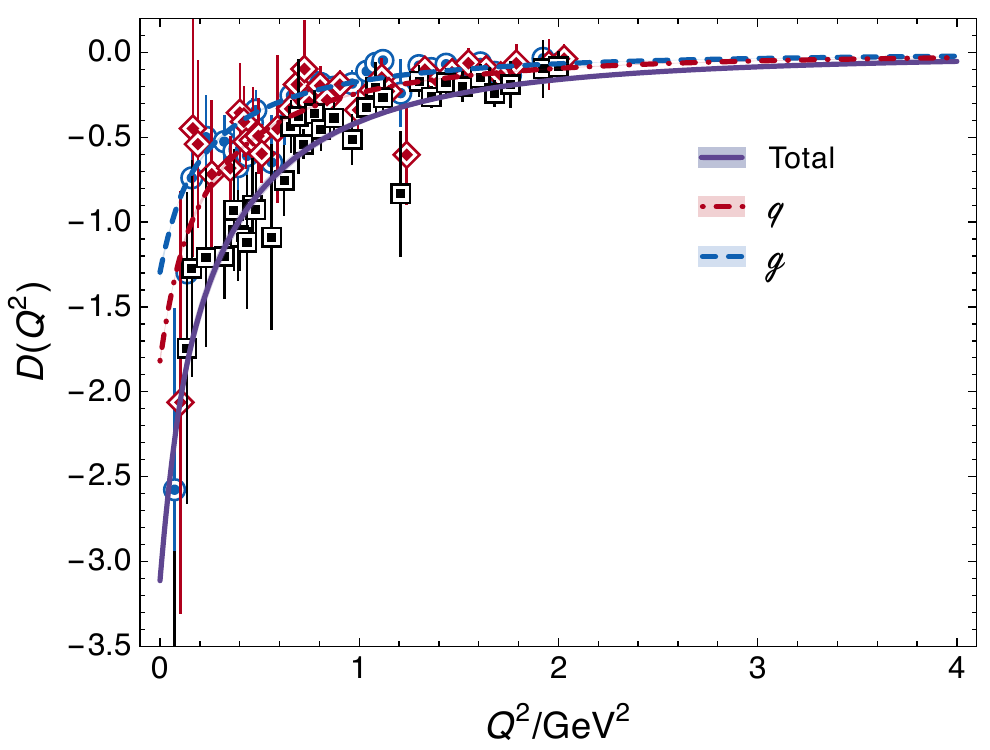}
\vspace*{-1.0ex}

\caption{\label{PlotGFFs}
Nucleon gravitational form factors.
Curves -- predictions herein: bracketing bands mark the extent of $1\sigma$ SPM uncertainty.
In each case, the overall (species-summed) result is independent of resolving scale, $\zeta$.
The species decompositions evolve with $\zeta$.
The lQCD points in each panel are reproduced from Ref.\,\cite{Hackett:2023rif}:
black squares -- total form factor;
blue circles -- glue component;
red diamonds -- quark.
}
\end{figure}

Our predictions for nucleon gravitational form factors are drawn in Fig.\,\ref{PlotGFFs}.  (Technical details describing the calculational procedures are provided in SupM-{\sf E}, which includes an explanation of the Schlessinger point method (SPM) used for extrapolation of analytic functions \cite{Schlessinger:1966zz, PhysRev.167.1411, Tripolt:2016cya, Cui:2021vgm}.)
%
%
The symmetry preserving character of our CSM analysis is evident in the values of $A(0)$, $J(0)$ -- see also Table~\ref{TableZero}.  Notably, in delivering $J(0)=1/2$, we confirm that the anomalous gravitomagnetic moment of a spin-$1/2$ system is zero \cite{Kobzarev:1962wt, Teryaev:1999su, Brodsky:2000ii}.  Furthermore, the value of the nucleon ``D-term'', \emph{viz}.\ $D(0)=-3.11(1)$ is a prediction.

\begin{table}[t]
\caption{\label{TableZero}
$Q^2=0$ values of proton gravitational form factors plus species decompositions at $\zeta = \zeta_2 = 2\,$GeV.
Our analysis enables a separation of light-quark valence (V) and sea (S) components and expresses a charm ($c$) quark sea contribution.
\emph{N.B}.\ Uncertainties on sea and glue components are typically anticorrelated with those on the valence components.  Such correlations are indicated by $\pm$, $\mp$.
lQCD results reproduced from Ref.\,\cite{Hackett:2023rif}: valence and sea are unseparated and no signal is reported for a $c$ contribution.
In both panels, $q$ is the sum over all quark contributions, valence and sea.
}
\begin{tabular*}
{\hsize}
{
l@{\extracolsep{0ptplus1fil}}
|l@{\extracolsep{0ptplus1fil}}
l@{\extracolsep{0ptplus1fil}}
l@{\extracolsep{0ptplus1fil}}}\hline\hline
herein$\ $ & $A(0)\ $ & $J(0)\ $ & $-D(0)\ $ \\\hline
Total$\ $ & $1.00\ $ & $0.50\ $ & $3.114(10)_\pm\ $ \\
$u_{\rm V}\ $ & $0.328(15)_\pm\ $ & $0.164(08)_\pm\ $ & $1.019(49)_\pm\ $ \\
$d_{\rm V}\ $ & $0.149(07)_\pm\ $ & $0.074(04)_\pm\ $ & $0.463(22)_\pm\ $ \\
$u_{\rm S}\ $ & $0.033(03)_\mp\ $ & $0.017(02)_\mp\ $ & $0.103(08)_\mp\ $ \\
$d_{\rm S}\ $ & $0.042(03)_\mp\ $ & $0.021(02)_\mp\ $ & $0.131(11)_\mp\ $ \\
$s\ $ & $0.025(02)_\mp\ $ & $0.012(02)_\mp\ $ & $0.077(06)_\mp\ $ \\
$c\ $ & $0.0086(05)_\mp\ $ & $0.0043(03)_\mp\ $ & $0.027(02)_\mp\ $ \\
$q\ $ & $0.584(13)_\pm\ $ & $0.292(06)_\pm\ $ & $1.820(43)_\pm\ $ \\
$g\ $ & $0.416(13)_\mp\ $ & $0.208(06)_\mp\ $ & $1.294(33)_\mp\ $ \\
%
\hline\hline
lQCD$\ $ & $A(0)\ $ & $J(0)\ $ & $-D(0)\ $ \\\hline
Total$\ $ & $1.011(37)\ $ & $0.506(25)\ $ & $3.87(97)\ $ \\
$u\ $ & $0.3255(92)\ $ & $0.2213(85)\ $ & $0.56(17)\ $ \\
$d\ $ & $0.1590(92)\ $ & $0.0197(85)\ $ & $0.57(17)\ $ \\
$s\ $ & $0.0257(95)\ $ & $0.0097(82)\ $ & $0.18(17)\ $ \\
$q\ $ & $0.510(25)\ $ & $0.251(21)\ $ & $1.30(49)\ $ \\
$g\ $ & $0.501(27)\ $ & $0.255(13)\ $ & $2.57(84)\ $ \\\hline\hline
\end{tabular*}
\end{table}

\smallskip

A species decomposition of the observable form factors is enabled by the all-orders (AO) evolution scheme detailed in Ref.\,\cite{Yin:2023dbw}.  Here we list its key tenets.
(\emph{a}) There is an effective charge, $\alpha_{1\ell}(k^2)$, in the sense of Refs.\,\cite{Grunberg:1980ja, Grunberg:1982fw} and reviewed in Ref.\,\cite{Deur:2023dzc}, that, when used to integrate the leading-order perturbative DGLAP equations \cite{Dokshitzer:1977sg, Gribov:1971zn, Lipatov:1974qm, Altarelli:1977zs}, defines an evolution scheme for \emph{every} parton distribution function (DF) that is all-orders exact.
The pointwise form of $\alpha_{1\ell}(k^2)$ is largely irrelevant.  Nevertheless, the process-independent strong running coupling defined and computed in Refs.\ \cite{Binosi:2016nme, Cui:2019dwv} has all requisite properties.
(\emph{b}) There is a scale, $\zeta_{\cal H}$, at which all properties of a given hadron are carried by its valence degrees-of-freedom.  At this scale, DFs associated with glue and sea quarks are zero.
Working with the charge discussed in Refs.\,\cite{Cui:2019dwv, Deur:2023dzc, Brodsky:2024zev}, the value of the hadron scale is a prediction \cite{Cui:2021mom}:
$\zeta_{\cal H} = 0.331(2)\,{\rm GeV}$.
Analysis of lQCD results relating to the pion valence quark DF yields a consistent result \cite{Lu:2023yna}: $\zeta_{\cal H} = 0.350(44)\,{\rm GeV}$.

Now consider a hadron $H$, some associated observable form factor $F(Q^2)$, and suppose one desires to reveal the contribution to $F(Q^2)$ from a parton species $\mathpzc p$ at resolving scale $\zeta$.
Using AO evolution, then for any quark or gluon sector contributions, $F^{\mathpzc q}(Q^2;\zeta)$, $F^{\mathpzc g}(Q^2;\zeta)$, respectively, which are expressed in terms of the first/zeroth Mellin moment of some combination of generalised parton distributions (GPDs) at zero skewness, $\xi=0$, the desired fractional contribution is $\langle x \rangle^{\mathpzc p}_\zeta \times F(Q^2)$ \cite[Sec.\,7, 8]{Raya:2021zrz}, where $\langle x \rangle^{\mathpzc p}_\zeta$ is the parton species light-front momentum fraction in $H$ at $\zeta$.
The nucleon gravitational form factors $A(Q^2)$, $J(Q^2)$ are in this class \cite{Polyakov:2018zvc}.
%
Following Ref.\,\cite[Secs.\,3.6\,-\,3.9]{Diehl:2003ny}, the AO scheme can be extended to obtain the same result for $D(Q^2)$.
Exploiting these features, we arrive at the $\zeta=\zeta_2:=2\,$GeV species decompositions drawn in Fig.\,\ref{PlotGFFs} and listed in Table~\ref{TableZero}.

\begin{figure}[t]
\centerline{%
\includegraphics[clip, width=0.44\textwidth]{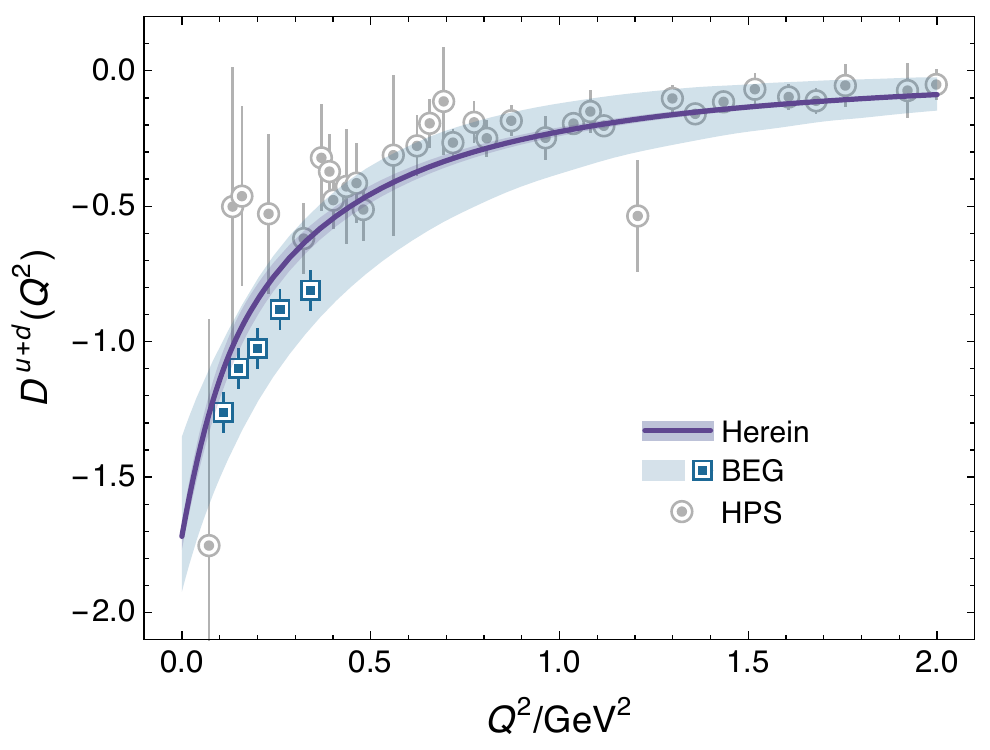}}
\caption{\label{FudDQ2}
Light-quark contribution to the proton $D$-term: $D^{u+d}(Q^2;\zeta_2)$.
Herein -- solid purple curve within like-coloured $1\sigma$ SPM uncertainty band;
BEG -- squares and associated fit from Ref.\,\cite{Burkert:2018bqq};
circles -- lQCD results from Ref.\,\cite{Hackett:2023rif}.}
\end{figure}

Considering light quarks alone, one finds $D^{u+d}(0;\zeta_2) = -1.73(5)$.
An inference from available data yields $D^{u+d}(0) = -1.63(29)$ \cite{Burkert:2018bqq}.
Moreover, notably, within uncertainties, our prediction is in $Q^2$ pointwise agreement with the extraction therein -- see Fig.\,\ref{FudDQ2}.

Figure~\ref{PlotGFFs} also compares our predictions for the species separated form factors with lQCD results.  Evidently, in all cases, they agree within mutual uncertainties.  One may quantify this by noting that the AO approach predicts that, for each form factor, the $\zeta_2$ contribution ratio gluon:total-quark is a fixed number (constant, independent of $Q^2$), \emph{viz}.\ ${\mathpzc g}(Q^2)/{\mathpzc q}(Q^2) = 0.71(4)$.  The lQCD results are consistent with this prediction -- see SupM-{\sf F}.  In fact, combining lQCD results for all form factors, one finds ${\mathpzc g}/{\mathpzc q} = 0.82(18)$, where the uncertainty marks $1\sigma$ about the central value.

Comparison of the panels in Table~\ref{TableZero} yields additional insights.
(\emph{a}) Considering the light-quark $Q^2=0$ contributions, our predictions are roughly in the ratio $u/d = 1.9$ for all form factors.  On the other hand, the lQCD results are: $A = 2.05(13)$; $J = 11.2(4.5)$; $D=0.98(42)$.  Namely, whilst the lQCD form factor with the smallest uncertainty, $A$, yields a ratio consistent with our prediction, the other two, with larger uncertainties, disagree greatly.
(\emph{b}) Turning to $A^{\mathpzc g}(0)$, unlike the lQCD result \cite{Hackett:2023nkr}, our prediction matches the value inferred from global analyses of data by many collaborations \cite{Hou:2019efy}: $0.413(6)$.
(\emph{c}) Our predicted values for both the quark and gluon contributions to $J(0)$ (proton spin) align with those reported in Refs.\,\cite{Cheng:2023kmt, Yu:2024qsd} (quark + diquark picture of the nucleon) and also
match the lQCD results reported in Ref.\,\cite{Alexandrou:2020sml}.
They differ from those in Ref.\,\cite{Hackett:2023nkr}: most notably, herein and in Refs.\,\cite{Cheng:2023kmt, Yu:2024qsd, Alexandrou:2020sml}, the quark contribution is greater than the glue contribution, whereas there is a signal for this being reversed in Ref.\,\cite{Hackett:2023nkr}.
(\emph{d})
Perhaps of most interest, we predict that the glue contribution to $D(0)$ is noticeably smaller than the quark contribution, with a ratio $0.71(4)$.  On the other hand, Ref.\,\cite{Hackett:2023nkr} reports a different result, with glue markedly greater than quark: $1.97(65)$, albeit with large uncertainty.

\smallskip

\begin{figure}[t]
\leftline{\hspace*{1.4ex}\includegraphics[clip, width=0.43\textwidth, height=0.3\textwidth]{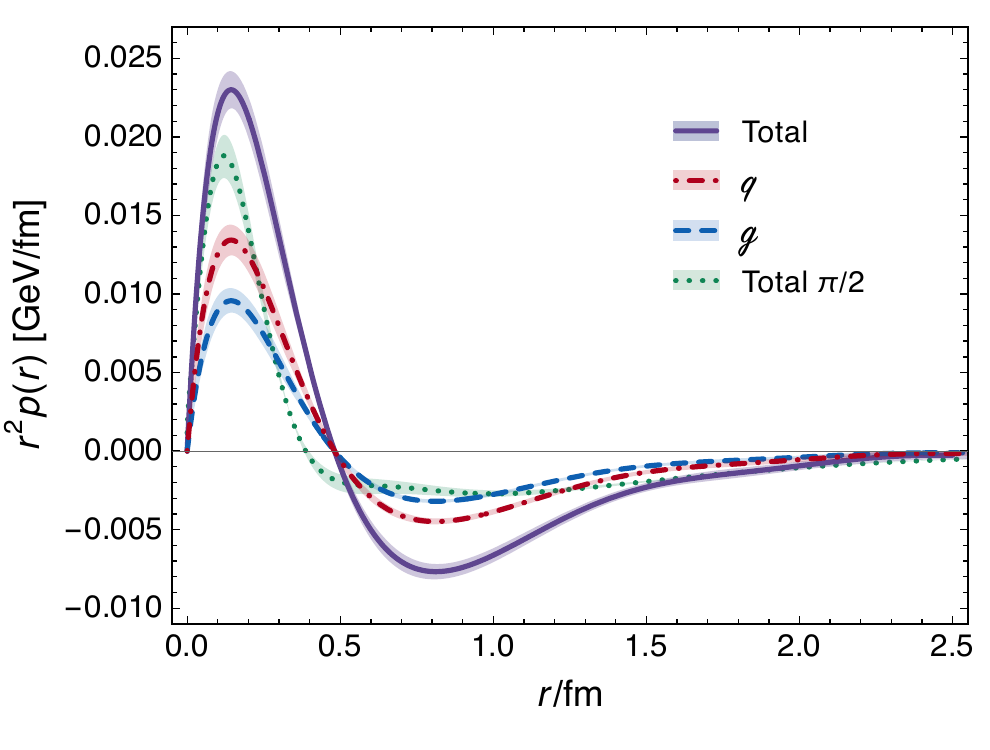}}
\vspace*{-39ex}

\leftline{\hspace*{0.5em}{\large{\textsf{A}}}}

\vspace*{39ex}

\leftline{\hspace*{0.5em}{\large{\textsf{B}}}}
\vspace*{-3ex}
\includegraphics[clip, width=0.41\textwidth]{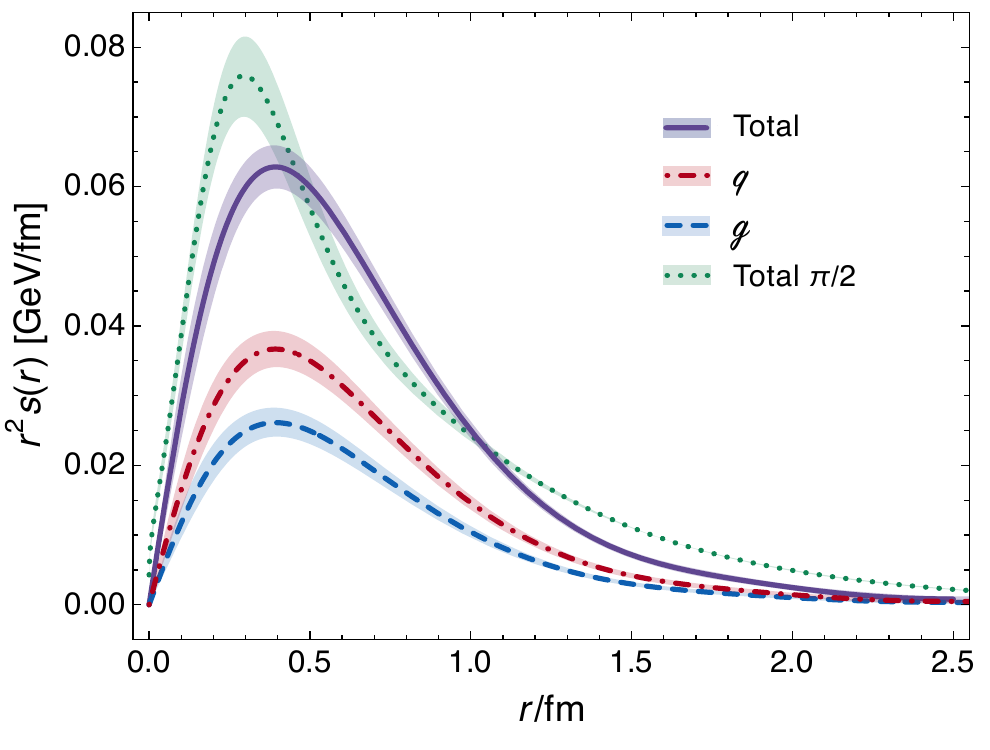}
\vspace*{4.5ex}
\caption{\label{PressureShear}
Nucleon pressure, $p(r)$, and shear force, $s(r)$, distributions, along with total-quark and glue species decompositions at $\zeta_2$.
Like pion distributions are drawn in green.  \emph{N.B}.\ The pion distributions are divided by $2$.
In all cases, the like-coloured band marks the extent of $1\sigma$ SPM uncertainty.
}
\end{figure}

\noindent\emph{5.\,Density Profiles}\,---\,%
Working with our Poincar\'e-invariant nucleon gravitational form factors, one may calculate an array of Breit frame density profiles (energy, pressure, shear and normal force distributions) via appropriate three-dimensional Fourier transforms \cite{Polyakov:2018zvc}.  (The relevant formulae are reproduced in SupM-{\sf G}.)
Figure\,\,\ref{PressureShear} displays our predictions for the nucleon pressure and shear force distributions in comparison with those in the pion, calculated using the same framework \cite{Xu:2023izo}.   Evidently, the pion peak values are roughly twice those in the proton.
It is here worth reiterating \cite{Burkert:2018bqq, Xu:2023izo} that such pressures are an order of magnitude greater than are expected at the core of neutron stars \cite{Ozel:2016oaf}.

It is worth discussing nucleon mass and mechanical radii.  The former is determined by  $A(Q^2)$, $D(Q^2)$ and the latter by $D(Q^2)$ alone (SupM-{\sf G}):
\begin{equation}
r_{\rm mass} = 0.81(5) r_{\rm ch}\,,\quad
r_{\rm mech} = 0.72(2) r_{\rm ch}\,,
\end{equation}
where $r_{\rm ch}=0.887(3)\,$fm is the proton charge radius calculated using the same framework \cite{Yao:2024uej}.  Regarding species decompositions, at $\zeta_2$:
$r_{\rm mass}^{\mathpzc q}=0.62(4) r_{\rm ch}$, $r_{\rm mass}^{\mathpzc g}=0.52(3) r_{\rm ch}$,
and
$r_{\rm mech}^{\mathpzc q}=0.55(2) r_{\rm ch}$, $r_{\rm mech}^{\mathpzc g}=0.47(2) r_{\rm ch}$.
Reviewing these predictions, we are led to stress the following points.
(\emph{a}) Our analysis is the only one available that provides a unified single-framework (phenomenological or theoretical) treatment of nucleon electromagnetic and gravitational form factors.  This makes the relative radii ordering statements unique and substantive.
(\emph{b}) The nucleon's mechanical radius is less than its mass radius.  
On the other hand, the available lQCD comparison suggests the reverse ordering, although the two radii are equal within lQCD uncertainties.
(\emph{c}) Alike with the pion \cite{Xu:2023bwv}, the proton's mass radius is less than its charge radius.
(\emph{d}) Regarding species separated radii at $\zeta_2$, we predict the quark contribution is greater than that from glue.  The reverse ordering is favoured in Ref.\,\cite{Hackett:2023nkr}.  In our view, considering Table~\ref{TableZero} and the associated discussion, this owes largely to the lQCD overestimate of the glue contribution to $D(Q^2)$.
(\emph{e}) Within uncertainties, our prediction for the quark pressure, Fig.\,\ref{PressureShear}A, matches that inferred from data \cite{Burkert:2018bqq}.  Likewise, our result for $r_{\rm mech}$ matches that inferred therein.
(\emph{f}) Considering the discussion in Refs.\,\cite{Du:2020bqj, Xu:2021mju, Sun:2021pyw, Tang:2024pky}, comparisons with currently reported inferences of glue contributions to proton gravitational form factors are unwarranted.

\smallskip

\noindent\emph{6.\,Summary}\,---\,%
%
Using a 
symmetry-preserving, 
sys\-te\-ma\-ti\-cally-improvable
truncation of the quantum field equations relevant to calculation of hadron properties, this study completes the first single-framework, unifying treatment of pion, kaon, nucleon electromagnetic and gravitational form factors.
Each element in the study is a well-defined approximation to an analogous quantity in quantum chromodynamics (QCD).
A single parameter characterises the quark + quark scattering kernel.  It is fixed with reference to gauge sector dynamics in QCD; so, all predictions are parameter-free and link observables directly with the three pillars of emergent hadron mass.

The analysis reveals the following features.
Regarding a separation of each of the nucleon's gravitational form factors (mass, spin, D-term) into glue and quark pieces, the ratio of these contributions is the same $Q^2$-independent number for all three: glue/quark $\approx 0.71$.  This prediction is consistent with available results from lattice-regularised QCD.
The near-core pressure in the pion is roughly twice that in the proton, so both are at least an order of magnitude greater than that of a neutron star.
The nucleon's mechanical radius is less than its mass radius, which in turn is less than its charge radius.
At a standard resolving scale, $\zeta=2\,$GeV, the glue contributions to both are 15\% less than the total quark contributions.
Within uncertainties, our predictions are consistent with phenomenological inferences from available data on deeply virtual Compton scattering.

The breadths of application and success of the parameter-free framework employed herein suggest that the predictions delivered should serve as benchmarks for future phenomenology and theory.  No improvement on this study can be anticipated before lattice-QCD analyses have simultaneously delivered infinite-volume, continuum-limit, carefully renormalised, physical pion mass studies of all physical properties discussed herein, \emph{viz}.\ pion, kaon, nucleon electromagnetic and gravitational form factors.

\smallskip

\noindent\emph{Acknowledgments}\,---\,%
We are grateful for constructive interactions with
S.-X.\ Qin. 
Work supported by:
National Natural Science Foundation of China (grant no.\ 12135007, 12233002);
Natural Science Foundation of Jiangsu Province (grant no.\ BK20220122);
Helmholtz-Zentrum Dresden-Rossendorf, under the High Potential Programme;
Spanish Ministry of Science and Innovation (MICINN grant no.\ PID2022-140440NB-C22);
and
Junta de Andaluc{\'{\i}}a (grant no.\ P18-FR-5057).

\medskip

\centerline{\rule{5em}{0.2ex}\;\textbf{Supplemental Material}\;\rule{5em}{0.2ex}}

\smallskip

\renewcommand{\theequation}{S.\arabic{equation}}
\renewcommand{\thefigure}{S.\arabic{figure}}

\begin{figure}[b]
\centerline{%
\includegraphics[clip, width=0.46\textwidth]{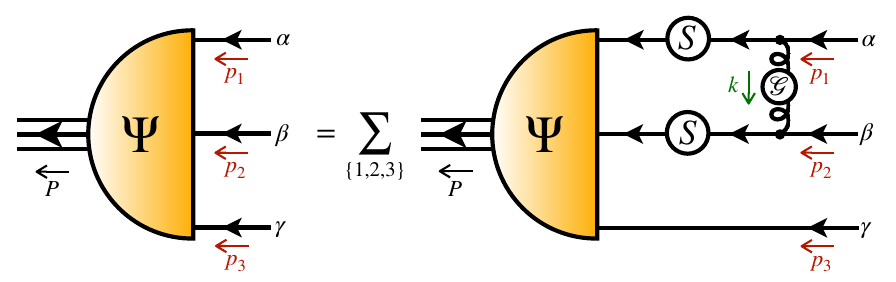}}
\caption{\label{FigFaddeev}
Faddeev equation. 
Filled semicircle: Faddeev amplitude, $\Psi$, the matrix-valued solution, which involves 128 independent scalar functions.
Spring: dressed-gluon interaction that mediates quark + quark scattering, Eq.\,(2).
Solid line: dressed-quark propagator, $S$, calculated from the rainbow gap equation.
Lines not adorned with a shaded circle are amputated.  Isospin symmetry is assumed.}
\end{figure}

\noindent{\sf A. Faddeev Equation}\,---\,The RL truncation nucleon Faddeev equation is drawn in Fig.\,\ref{FigFaddeev}.  It is constructed using the Bethe-Salpeter kernel -- Eq.\,(2) -- and the dressed light-quark propagator, which has the general form
\begin{equation}
S(k) = 1/[ i \gamma\cdot k A(k^2) + \mathbf 1 B(k^2)]\,.
\end{equation}
Discussions of the formulation and solution of this linear, homogeneous integral equation are provided, \emph{e.g}., in Refs.\,\cite{Qin:2018dqp, Wang:2018kto, Qin:2019hgk}.  The output is the nucleon Faddeev amplitude, which can be used to compute all form factor matrix elements, electromagnetic, weak, gravitational, etc.  In any such calculation, the amplitude must be normalised.  The canonically normalised amplitude ensures, \emph{e.g}., that the proton electric charge is unity.

\medskip

\begin{figure}[t]
\centerline{%
\includegraphics[clip, width=0.46\textwidth]{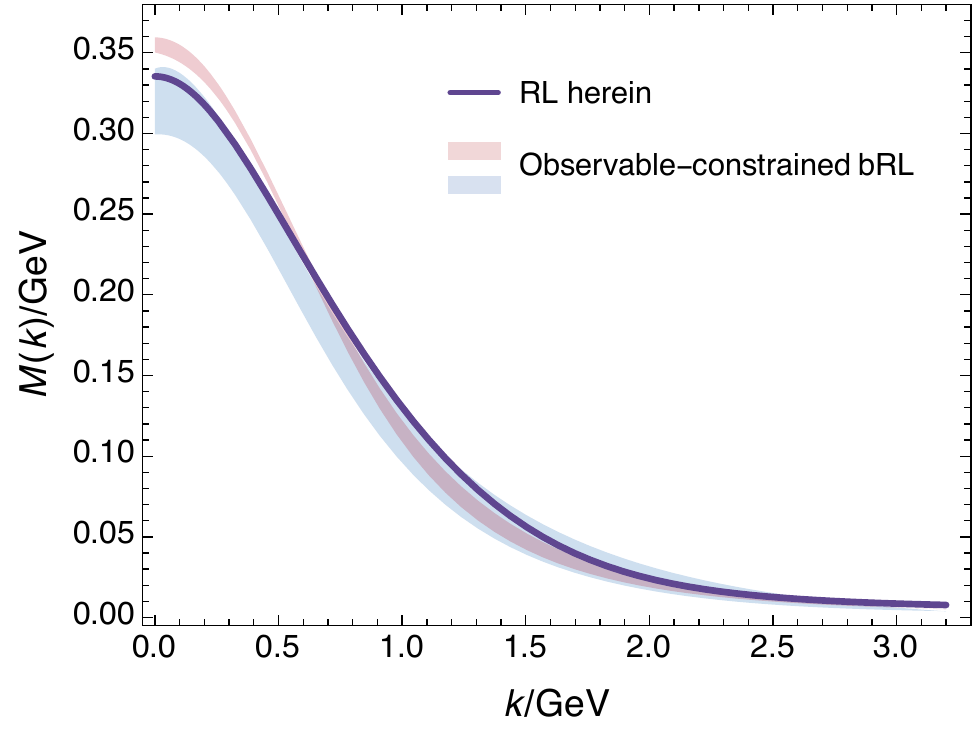}}
\caption{\label{FigMk}
Dressed light quark mass function.
Solid purple curve -- RL result calculated and used herein to deliver all predictions.
Blue and red bands -- range of results admitted by the observable-constrained, nonperturbative beyond-RL (bRL) truncation described in Ref.\,\cite{Binosi:2016wcx}.
}
\end{figure}

\begin{figure*}[t]
\centerline{%
\includegraphics[clip, width=0.85\textwidth]{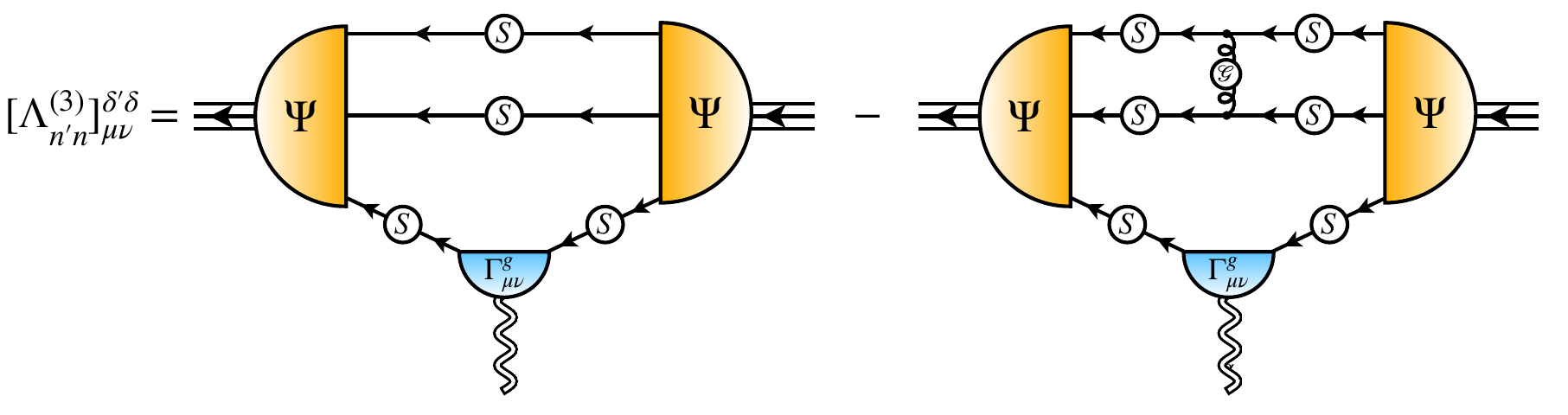}}
\caption{\label{FigCurrent}
$a=3$ spinor component of the nucleon current associated with the energy-momentum tensor, where
$\delta$, $\delta^\prime$ are spinor indices and $n$, $n^\prime$ are isospin indices.
${\mathpzc G}$ indicates the Bethe-Salpeter kernel, also used to define the RL truncation gap equation that yields the dressed light-quark propagator, $S$, as its solution.
Together, these elements complete the kernel in the Faddeev equation that yields $\Psi$, the nucleon Faddeev amplitude -- see Fig.\,S.1.
The remaining element is $\Gamma_{\mu\nu}^g$, the dressed-graviton+quark vertex, obtained following Ref.\,\cite{Xu:2023izo}.
The complete current has three terms: $\Lambda_{\mu\nu}(Q) = \sum_{a=1,2,3} \Lambda_{\mu\nu}^a(Q)$; but using symmetries, one can readily obtain the $a=1,2$ components of the current once $a=3$ is known \cite[Appendix~B]{Eichmann:2011pv}.
}
\end{figure*}

\noindent{\sf B. Quark + Quark Interaction}\,---\,
It is here worth providing context for the interaction in
Eq.\,(3)
by noting that, following Ref.\,\cite{Qin:2011dd}, one may draw a connection between $\tilde{\mathpzc G}$
and QCD's process-independent effective charge, discussed in Refs.\,\cite{Cui:2019dwv, Deur:2023dzc, Brodsky:2024zev}.
That effective charge is characterised by an infrared coupling value $\hat\alpha(0)/\pi = 0.97(4)$ and a gluon mass-scale $\hat m_0 = 0.43(1)\,$GeV determined in a combined continuum and lattice analysis of QCD's gauge sector \cite{Cui:2019dwv}.
The following values are those of analogous quantities inferred from Eq.\,(2): 
\begin{equation}
\label{DiscussInteraction}
\alpha_{\mathpzc G}(0)/\pi = 1.45\,,\quad m_{\mathpzc G} = 0.54\,{\rm GeV}\,.
\end{equation}
They are a fair match with the QCD values, especially since earlier, less well informed versions of the RL interaction yielded $\alpha_{\mathpzc G}(0)/\pi \approx 15$, \emph{i.e}., a value ten-times larger~\cite{Qin:2011dd}.

Existing analyses of hadron properties suggest that inclusion of corrections to RL truncation in ground state channels has little impact beyond a relaxation of the numbers in Eq.\,\eqref{DiscussInteraction} toward their QCD values; for instance, compare the RL and EHM-improved spectra in Refs.\,\cite{Xu:2022kng}.  
This is further highlighted by Fig.\,\ref{FigMk}, which displays the mass function used herein to deliver the nucleon gravitational form factors.  
Obtained by solving the quark gap equation with the RL kernel specified by Eqs.\,\eqref{EqRLInteraction}, \eqref{defcalG}, it plainly falls comfortably within the solution range determined by solving the gap equation with the symmetry-preserving, nonperturbative beyond-RL truncation described in Ref.\,\cite{Binosi:2016wcx}.
It is for these reasons that the modern formulation of RL truncation provides a reliable, predictive tool -- see also, \emph{e.g}., Refs.\,\cite{Yao:2024drm, Yao:2024uej}, in which it was used to deliver pion, kaon, and nucleon electromagnetic form factors, and Ref.\,\cite{Xu:2023izo}, wherein pion and kaon gravitational form factors are calculated.

\medskip

\noindent{\sf C. Nucleon Gravitational Current}\,---\,
Generalising the electromagnetic current in Ref.\,\cite{Eichmann:2011vu} to the gravitational interaction is straightforward -- see Fig.\,\ref{FigCurrent}.

\medskip

\noindent{\sf D. Quark + Graviton Vertex}\,---\,
Here we briefly review the analysis in Ref.\,\cite{Xu:2023izo} -- those seeking details may refer therein.

The dressed graviton + quark vertex satisfies the following Ward-Green-Takahashi (WGT) identity:
\begin{equation}
\label{GvertexWGTI}
Q_\mu i\Gamma_{\mu\nu}^g(k,Q) =
S^{-1}(k_+) k_{- \nu} - S^{-1}(k_-) k_{+\nu}\,.
\end{equation}

In RL truncation, $\Gamma_{\mu\nu}^g$ is obtained by solving the following inhomogeneous Bethe-Salpeter equation:
\begin{align}
i \Gamma_{\mu\nu}^g&(k_+,k_-)  =
Z_2 [i \gamma_\mu k_\nu
- \delta_{\mu\nu} (i\gamma\cdot k+Z_m^0 m^\zeta)] \nonumber \\
& + Z_2^2 \int_{dl}^{\Lambda} \mathscr{K}(k-l) [ S (l_+) i \Gamma_{\mu\nu}^g(l_+,l_-) S (l_-)] \,,
\label{RLGQV}
\end{align}
where $k_{\pm} = k \pm Q/2$, $Z_m^0$ is the chiral-limit mass renormalisation constant, and $Z_2$ is the quark wave function renormalisation constant.
(In all analyses discussed herein, a mass-independent momentum-subtraction renormalisation scheme is employed \cite{Chang:2008ec}.)

With complete generality, the solution of Eq.\,\eqref{RLGQV} may be written thus:
\begin{equation}
\label{EqGQV}
\Gamma_{\mu\nu}^g(k,Q) = \Gamma_{\mu\nu}^{g_M}(k,Q) + \Gamma_{\mu\nu}^{gT}(k,Q)\,.
\end{equation}
The first term:
\begin{align}
i\Gamma_{\mu\nu}^{g_M}&(k,Q)  \nonumber \\
& = i\Gamma_\mu^{\rm BC}(k,Q) k_\nu
-\tfrac{1}{2}\delta_{\mu\nu}[ S^{-1}(k_+) + S^{-1}(k_-)]  \nonumber \\
&
\quad +  i T_{\mu\alpha}(Q)T_{\nu\beta}(Q)4 \hat{\Gamma}^2_{\alpha\beta}(k_+,k_-)\,, \label{GvertexWGTI2}
\end{align}
with
\begin{align}
i \Gamma_\nu^{\rm BC}(k_+,k_-)
&  = i \gamma_\nu \Sigma_{A_\pm} \nonumber \\
 & \quad + 2 i k_\nu \gamma\cdot k \, \Delta_{A_\pm} + 2 k_\nu \Delta_{B_\pm} \,,
 \label{BCAnsatz}
 %
\end{align}
$\Sigma_{A_\pm} = [A(k_+^2)+A(k_-^2)]/2$,
$\Delta_{F_\pm} = [F(k_+^2)-F(k_-^2)]/[k_+^2-k_-^2]$, $F=A,B$,
resolves Eq.\,\eqref{GvertexWGTI}.

${\Gamma}^2_{\alpha\beta}(k;Q)$ in Eq.\,\eqref{GvertexWGTI2} is the tensor + quark vertex generated by the inhomogeneity
\begin{equation}
\label{eq:G20}
\Gamma^{2}_{0\mu\nu}(k;Q) = T_{\mu\alpha}(Q)T_{\nu\beta}(Q) \tfrac{1}{2} \left(\gamma_\alpha k_\beta + \gamma_\beta k_\alpha \right) \,,
\end{equation}
with $\hat\Gamma(k , Q) = \Gamma(k , Q) - \Gamma(k , 0)$ so as to ensure the absence of kinematic singularities.
Dynamical singularities nevertheless appear in $\Gamma^{2}_{\mu\nu}$; namely, one at the pole position of each $I=0$ tensor meson.
Since ${\Gamma}^2_{\mu\nu}$ possesses eight independent Dirac matrix valued tensor structures, then Eq.\,\eqref{GvertexWGTI2} involves thirteen such nonzero terms.

The remaining term in Eq.\,\eqref{EqGQV}, $\Gamma_{\mu\nu}^{gT}(k,Q)$, satisfies $Q_\mu \Gamma_{\mu\nu}^{gT}(k,Q) = 0$.  It represents all possible transverse structures not already included in the first term.  Like $\Gamma^{2}_{\mu\nu}$, $\Gamma_{\mu\nu}^{gT}$ does not contribute to resolving Eq.\,\eqref{GvertexWGTI}.  It is obtained by solving the appropriate Bethe-Salpeter equation.  In doing so, one sees the emergence of isoscalar scalar mesons in the graviton + quark vertex.

As stressed in Ref.\,\cite{Xu:2023izo}, a few contributions dominate the graviton + light-quark vertex, \emph{viz}.\ those parts which saturate the WGT identity -- already included; pieces associated with an $f_2$ tensor meson pole -- incorporated via  ${\Gamma}^2_{\alpha\beta}(k;Q)$; and those tied to an analogous scalar meson resonance.  The last is captured by writing
\begin{equation}
 \Gamma_{\mu\nu}^{gT}(k,Q) =  T_{\mu\nu}(Q) \Gamma_{\mathbb I}(k;Q)\,,
 \label{GQVSC}
\end{equation}
where $\Gamma_{\mathbb I}(k;Q)$, which has four independent Dirac matrix valued structures, $D_{\mathbb I}^{j=1,4}\propto\{\mathbf 1, \gamma\cdot k, \gamma\cdot Q, \sigma_{\alpha\beta}k_\alpha Q_\beta\}$ -- see, \emph{e.g}., Ref.\,\cite[Appendix~A]{Krassnigg:2009zh}, is obtained by solving
\begin{align}
& {\rm tr}_D{\mathpzc P}_{\mu\nu}^j(k,Q)\Gamma_{\mu\nu}^{gT}(k_+k_-) \nonumber \\
=
& {\rm tr}_D {\mathpzc P}_{\mu\nu}^j(k,Q) Z_2^2 \int_{dl}^{\Lambda} \mathscr{K}(k-l) S (l_+) \nonumber \\
& \quad \times
 \{ \Gamma_{\mu\nu}^{g_M}(l_+,l_-) + T_{\mu\nu}(Q) \Gamma_{\mathbb I}(l_+;l_-)  \}  S (l_-) \,.
\label{GammaIRL}
\end{align}
Here,
\begin{subequations}
\begin{align}
{\rm tr}{\mathpzc P}_{\mu\nu}^j(k,Q) \Gamma_{\mu\nu}^{gT}(k,Q) & = D_{\mathbb I}^{j},\\
{\rm tr}{\mathpzc P}_{\mu\nu}^j(k,Q) \Gamma_{\mu\nu}^{g_M}(k,Q) & \equiv 0 \,.
\end{align}
\end{subequations}

At this point, gathering all terms described above, one has an efficacious solution for the graviton + quark vertex.  It involves $2$ one-variable and $12$ two-variable functions and various distinct, associated Dirac matrix structures.
For completeness, we list our results for the lightest scalar (${\mathbb S}$) and tensor (${\mathbb T}$) mesons, \emph{viz}.\
the calculated masses and residue coefficients (in GeV):
\begin{equation}
\label{PoleValues}
\begin{array}{c|cccc}
    & m_{\mathbb S} & f_{\mathbb S} & m_{\mathbb T} & f_{\mathbb T} \\ \hline
%
%
u=d & 0.53 & 0.025 & 1.20 & 0.042 \\
%
\end{array}\,.
\end{equation}

\medskip

\noindent{\sf E. Form Factor Calculation and SPM}\,---\,
%
The Faddeev amplitude depends on the nucleon total momentum, $P$, and two relative momenta, $p$, $q$; so each function in the amplitude depends on three angular variables defined via the inner products $p\cdot q$, $p\cdot P$, $q\cdot P$.
In solving the Faddeev equation, we used eight Chebyshev polynomials to express the dependence on each angle \cite{Maris:1997tm}.
This enables evaluation of $\Psi$ at any required integration point in either the Faddeev equation or the current.
$P$ is a complex-valued (timelike) vector, $P^2=-m_N^2$, whereas $Q$ is spacelike.
Thus, when evaluating the current, the integrand sample points are typically in the complex plane and the integrand exhibits oscillations whose amplitudes grow with $Q^2$ \cite{Maris:2000sk}.
To handle this, increasing the number of Chebyshev polynomials and quadrature points is effective on $Q^2 \leq Q_{\rm m}^2$, $Q_{\rm m}^2 \approx 2.5\,$GeV$^2$.
At larger $Q^2$ values, however, such a brute force approach fails to deliver accurate results.
(Regarding nucleon electromagnetic form factors, where $\Gamma_{\mu\nu}^g$ is replaced by the simpler photon + quark vertex, direct reach to larger $Q^2$ is possible \cite{Yao:2024uej}.)

To obtain results on $Q^2\gtrsim 2.5\,$GeV$^2$, we extrapolate using the Schlessinger point method (SPM) \cite{Schlessinger:1966zz, PhysRev.167.1411, Tripolt:2016cya}.
The SPM is grounded in analytic function theory and based on the Pad\'e approximant.
It accurately reconstructs any function in the complex plane within a radius of convergence determined by that one of the function's branch points which lies closest to the real domain that provides the sample points.  Modern implementations introduce a statistical element; so, the extrapolations come with an objective and reliable quantitative estimate of uncertainty.
Crucially, the SPM is free from practitioner-induced bias; so, delivers objective analytic continuations.
In practice, the SPM has been blind-tested against numerous models and physically validated in applications that include
determination of hadron and light nucleus radii from electron scattering \cite{Cui:2022fyr};
extraction of resonance properties from scattering data \cite{Binosi:2022ydc};
searching for evidence of the odderon in high-energy elastic hadron+hadron scattering \cite{Cui:2022dcm};
and
calculation of meson and baryon electromagnetic form factors \cite{Yao:2024drm, Yao:2024uej}.
%

Our SPM extrapolations are developed as follows.
\begin{description}
\item[Step 1]
For each form factor, we produce $N=30$ directly calculated values of $Q^2 \times\,$form factor, spaced evenly on $Q^2\lesssim 2.5\,$GeV$^2$.
\item[Step 2]
From that set, $M_0=6$ points are chosen at random,
the usual SPM continued fraction interpolation is constructed,
and that function is extrapolated onto $Q^2 > Q_{\rm m}^2$.
The curve is retained so long as it is singularity free on $Q^2 \leq 100\,$GeV$^2$.
\item[Step 3]
Step~2 is repeated with another set of $M_0$ randomly chosen points.
\item[Step 4]
One continues with 2 and 3 until $n_{M_0}=200\,$ smooth extrapolations are obtained.
\item[Step 5]
Steps 2 and 3 are repeated for $M=\{M_0+ 2 i | i=1,\ldots,5\}$.
\item [Step 6]
At this point, one has $1\,200$ statistically independent extrapolations for $Q^2 \times\,$form factor.
\end{description}
Working with these extrapolations, then at each value of $Q^2$, we record the mean value of all curves as the central prediction and report as the uncertainty the function range which contains 68\% of all the extrapolations -- this is a $1\sigma$ band.

\medskip

\begin{figure}[t]
\vspace*{0.2em}

\leftline{\hspace*{0.5em}{\large{\textsf{A}}}}
\vspace*{-3ex}
\includegraphics[clip, width=0.41\textwidth]{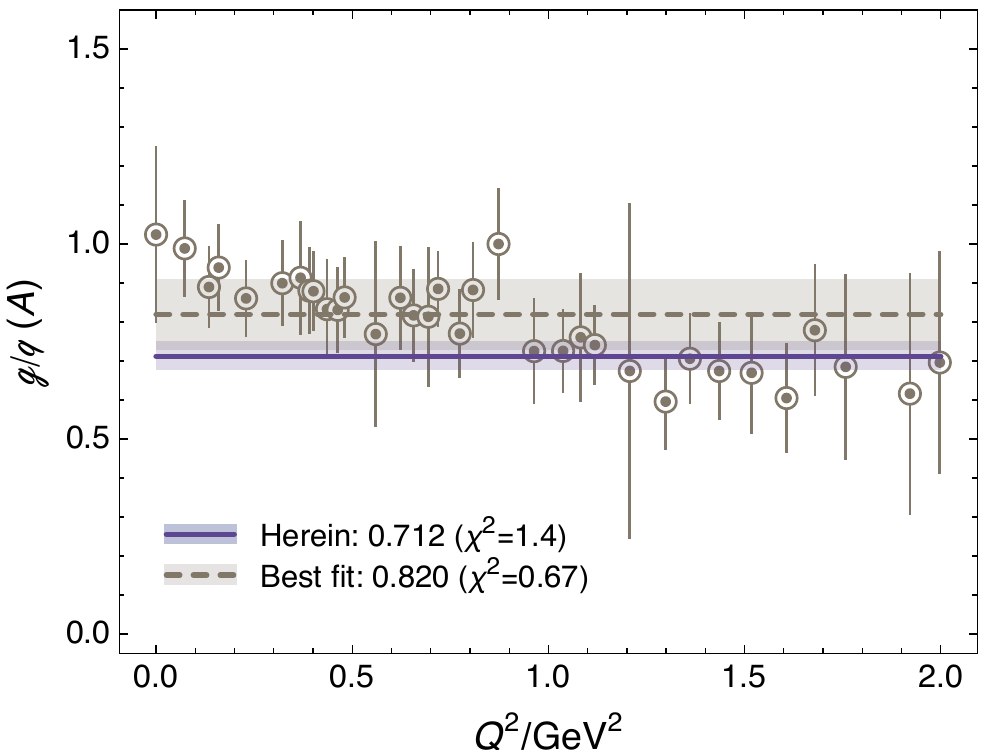}
\vspace*{2ex}

\leftline{\hspace*{0.5em}{\large{\textsf{B}}}}
\vspace*{-2ex}
\includegraphics[clip, width=0.41\textwidth]{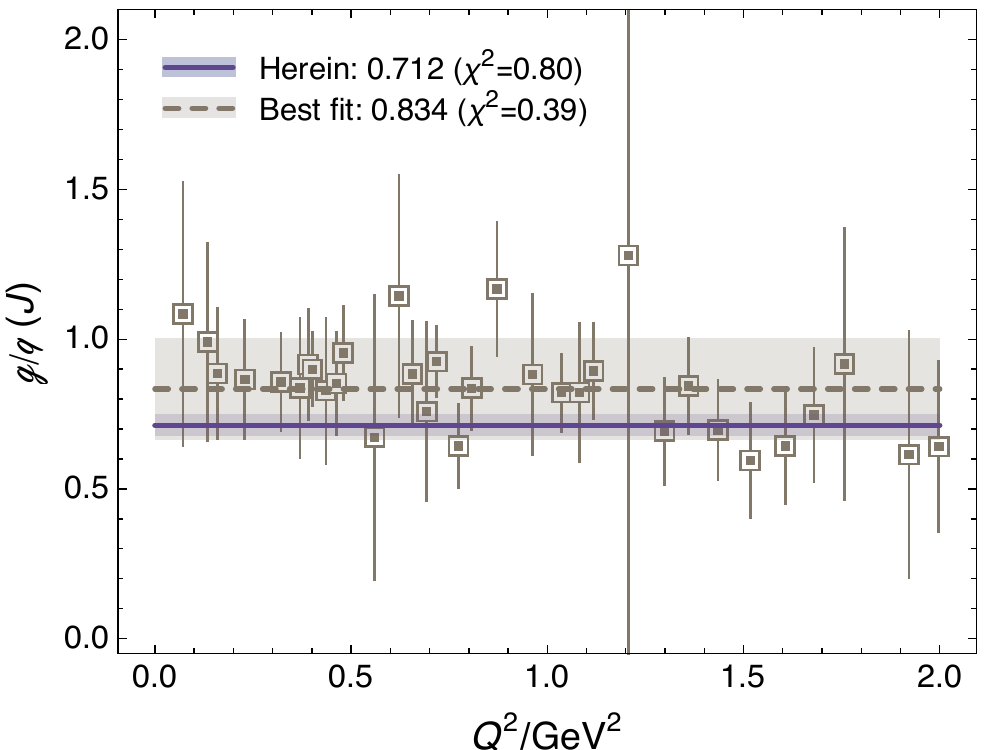}
\vspace*{1.7ex}

\leftline{\hspace*{0.5em}{\large{\textsf{C}}}}
\vspace*{-2ex}
\includegraphics[clip, width=0.41\textwidth]{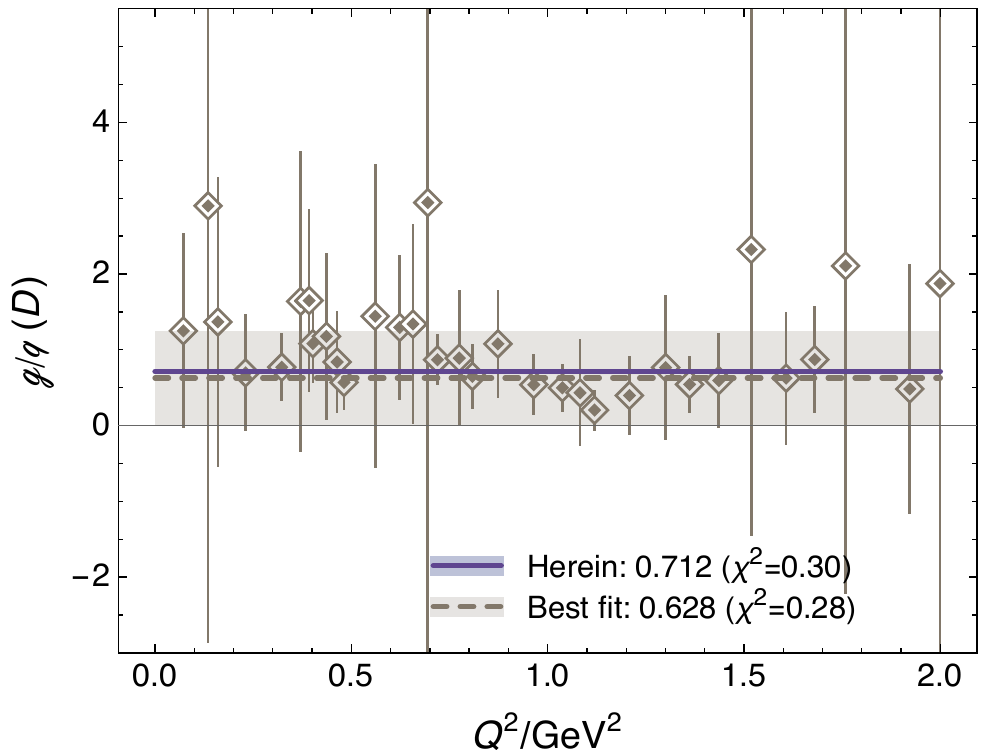}
\vspace*{1.5ex}

\caption{\label{gqratio}
Ratio of glue:quark-singlet form factor contributions.
Our prediction: $0.71(4)$ for each form factor -- solid purple curve and band.
Points (each panel) -- lQCD results \cite{Hackett:2023rif} along with the associated linear least-squares fit and bracketing band that extends to $1\sigma$ around the central value.
}
\end{figure}

\begin{figure}[t]
\centerline{%
\includegraphics[clip, width=0.42\textwidth]{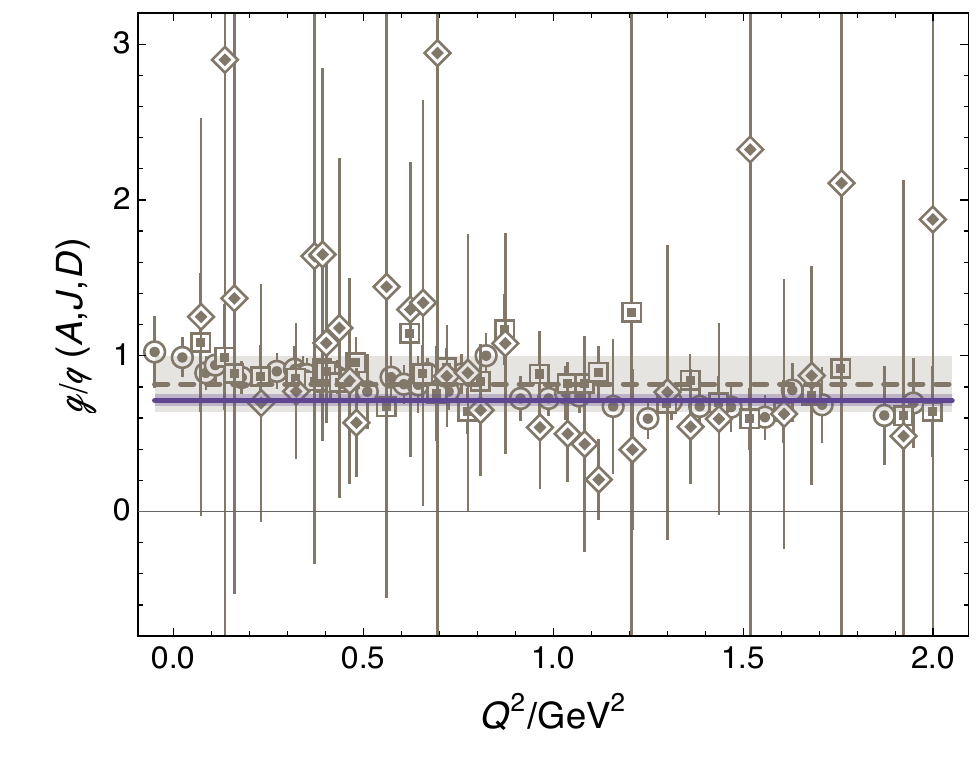}}

\vspace*{-2ex}

\caption{\label{gqratioAll}
Combined $A$, $J$, $D$ lQCD ${\mathpzc g}/{\mathpzc q}$ results (grey points): grey line -- uncertainty weighted average of all points; and grey band -- $1\sigma$ around the central value: $0.82(18)$.
Prediction herein: ${\mathpzc g}/{\mathpzc q}=0.71(4) $ (purple line and band).
}
\end{figure}

\noindent{\sf F. Parton Species Decompositions}\,---\,
%
Regarding species decompositions, the AO scheme predicts the following \cite{Yu:2024qsd}:
\begin{equation}
\label{EqRatio}
F=A, J, D \; | \; \frac{F^{\mathpzc g}(Q^2)}{\sum_{\mathpzc q} F^{\mathpzc q}(Q^2)} = 0.71(4)\,.
\end{equation}
This prediction is compared with lQCD results \cite{Hackett:2023rif} in Figs.\,\ref{gqratio}, \ref{gqratioAll}.
The uncertainty on the lQCD results is large.
Nevertheless, they are compatible with Eq.\,\eqref{EqRatio}; namely, for each form factor, the lQCD results are (\emph{a}) consistent with a $Q^2$-independent ratio of glue-to-quark contributions and (\emph{b}) a value of this ratio that matches our prediction within uncertainties.

\medskip

\noindent{\sf G. Breit Frame Density Profiles}\,---\,Energy, pressure, and shear force distributions have been defined via the following formulae \cite{Polyakov:2018zvc} ($t=Q^2$):
\begin{subequations}
\label{profilest}
\begin{align}
\epsilon(r) & = m_N ( \hat A(r) - \tfrac{1}{4m_N^2}[\widehat{(t D)}(r) + \widehat{(t A)}(r) - 2 \widehat{(t J)}(r)]), \\
p(r) & = \frac{1}{6m_N}\frac{1}{r^2} \frac{d}{dr} r^2 \frac{d}{dr} \hat D(r) \,, \\
s(r) & = -\frac{1}{4m_N}r \frac{d}{dr} \frac{1}{r} \frac{d}{dr} \hat D(r)\,,
\end{align}
\end{subequations}
which involve the Fourier transform
\begin{equation}
  \hat f(|\vec{r}|) = \int \frac{d^3 q}{(2\pi)^3}{\rm e}^{\vec{q}\cdot \vec{r}} f(t \to |\vec{q}|^2)\,.
\end{equation}
(Species decompositions may be obtained by replacing each total form factor by its species component.)
It is proper to note that questions concerning the interpretation of such distributions have been widely canvassed -- see, \emph{e.g}., Refs.\,\cite{Jaffe:2020ebz, Freese:2021czn, Freese:2021mzg, Epelbaum:2022fjc, Freese:2022fat}.  Other transforms are possible \cite{Freese:2022fat}.
Nevertheless, since the input function in any case is always the same, then no projective mapping, like the construction of a density in two or three spacelike dimensions, can deliver any objective information that is not already contained in the Poincar\'e-invariant subject function.
Thus, whatever type of transform is chosen, it is merely a mathematical operation on the same input object; hence, interpreted judiciously, all outputs are qualitatively equi\-valent.

In terms of the quantities just defined, the normal force distribution in the nucleon is
\begin{equation}
F^\parallel (r) = p(r) + (2/3) s(r)\,.
\end{equation}
Nucleon mass and mechanical radii can now be defined in terms of $\epsilon(r)$, $F^\parallel (r)$ \cite{Polyakov:2018zvc}:
\begin{subequations}
\label{defradii}
\begin{align}
\langle r^2 \rangle_{\rm mass} & = \frac{\int d^3 r \, r^2 \epsilon(r)}{\int d^3 r  \epsilon(r)}\,, \\
\langle r^2 \rangle_{\rm mech} & = \frac{\int d^3 r \, r^2 F^\parallel(r)}{\int d^3 r  F^\parallel(r)}\,.
\end{align}
\end{subequations}
In a common interpretation, the mechanical radius measures the extent of the nucleon's normal force distribution.
These expressions are equivalent to the following:
\begin{subequations}
\begin{align}
\langle r^2 \rangle_{\rm mass} & = \left[\left. -6 \frac{d}{dt} A(t) \right|_{t=0} - 3 \frac{D(0)}{2 m_N^2} \right]\frac{1}{A(0)}\,, \label{massradius}\\
\langle r^2 \rangle_{\rm mech} & =  \frac{6 }{\int_0^\infty dt \, [D(t)/D(0)]}\,,
\label{mechradius}
\end{align}
\end{subequations}
which are typically easier to use. 
Consider Eq.\,\eqref{massradius}.
(\emph{a}) Regarding the total mass form factor, $A(0)=1$; but this is not true for species separated contributions.
(\emph{b}) The second term is large and positive: using our predictions, the value is $(0.45\,{\rm fm})^2$.
Turning to Eq.\,\eqref{mechradius}, it is evident that the mechanical radius is
determined solely by the integral of the $Q^2=0$ unit-normalised version of the nucleon $D$-term function.


\end{document}